\documentstyle[12pt]{article}
\def\no{\nonumber}
\begin{document}

\begin{titlepage}

\begin{flushright}

ULB-TH--98/05

\end{flushright}

%\hfill 
\vspace{1cm}

\begin{centering}

{\Large \bf 
Anomalous Global Currents and}

\vspace{.5cm}

{\Large \bf Compensating Fields in the BV Formalism}
 
\vspace{1cm}

{\large Ricardo Amorim$^{a}$, Nelson R. F. Braga$^{a}$
and Marc Henneaux$^{b,c }$}

\vspace{0.5cm}

{\sl
$^a$ Instituto de F\'{\i}sica, Universidade Federal
do Rio de Janeiro,\\
Caixa Postal 68528, 21945-970  Rio de Janeiro, Brazil\\[1.5ex]

$^b$ Facult\'e des Sciences, Universit\'e Libre de
Bruxelles,\\
Campus Plaine C.P. 231, B--1050 Bruxelles, Belgium\\[1.5ex]

$^c$ Centro de Estudios Cient\'\i ficos de Santiago,\\
Casilla 16443, Santiago 9, Chile}

\vspace{1cm}

\begin{abstract}
We compute the anomalous divergence of currents associated with global
transformations in the antifield formalism, by introducing
compensating fields that gauge these transformations.  We
consider the explicit case of the global axial current in QCD but
the method applies to any global
transformation of the fields.
\end{abstract}

\end{centering}

\vspace{1cm}


\vfill

\noindent{\tt amorim@if.ufrj.br;
braga@if.ufrj.br; henneaux@ulb.ac.be}
\end{titlepage}

\pagebreak

\section{Introduction}
\setcounter{equation}{0}

It is well known that quantum corrections can modify 
the expectation values of the divergence of global currents \cite{ABJ,Ja}.
In particular, a classically vanishing divergence of a global current can acquire 
a non vanishing expectation value at the quantum level.
It was shown by Fujikawa \cite{Fuj} that these quantum contributions 
can be calculated by path integral methods if one appropriately 
regularizes the functional measure.

The Batalin Vilkovisky (BV), or field-antifield formalism, is an 
extremely powerful procedure for the  quantization of gauge theories
\cite {ZJ0,BV1,HT}.  The occurrence of local anomalies in this
formalism has been discussed in \cite{TPN}, where they have 
been related to the non-invariance of the measure under (rigid) BRST
transformations.  The purpose of this letter
is to develop a method for computing the anomalous divergence
of currents associated with {\em global} (as
opposed to local) transformations.  To that end, we 
introduce pure gauge ``compensating fields" \cite{CF} that couple to
the divergence in question.  We then turn to the master
equation and show that quantum corrections are needed in
order to fulfill the quantum master equation.  These quantum
corrections to the solution of the master equation do exist (no gauge anomaly)
and turn out to be crucial for our purposes.  Indeed, they
precisely generate the quantum corrections
to the divergence of the global current. This is easily seen
by choosing appropriately
the gauge for the new gauge freedom
and using the standard Fradkin-Vilkovisky theorem of the antifield formalism.
[Rigid symmetries have been discussed from a different
point of view in the antifield formalism in \cite{BHW}.  See
also \cite{HSken}].

In reference \cite{AB} a procedure for calculating anomalous divergences in the particular case of Abelian global symmetries was presented.
In that paper  the original symmetry content of the action was enlarged by the introduction of extra gauge fields that could be trivially removed by a gauge fixing.
At the quantum level the symmetries introduced in that way are apparently broken but can be trivially restored with the introduction of appropriate counterterms.
This could be done because of the cohomological triviality of the field extension, which was easily proved once the new fields and the corresponding ghosts are combined in BRST doublets.
The anomalous divergences of the currents associated with the enlarged symmetries are then calculated by using the independence of the path integral with respect to the gauge fixing. 

In the present paper the treatment of \cite{AB} is generalized in several non trivial ways. First, we are considering non Abelian global transformations in the context of an original theory that presents itself a non Abelian local symmetry. The gauging pr
ocedure,
necessary for the calculation of the anomalous current divergences, mix non-trivially both kinds of symmetries. We show in section (2) that the resulting gauge structure becomes actually a semi-direct product of $SU(N)$ with itself instead of a direct pro
duct.
This non trivial algebraic structure reflects itself in the process of quantization.
For instance, the cohomological triviality of the extension can  only be proved in a much more elaborated way (see section (4) ) since the new fields only form  BRST doublets in the Abelian limit. Also, the form of the counterterms is not a trivial genera
lization of the one of reference \cite{AB} but relies heavily on peculiar aspects of the Lie algebra cohomology of non-abelian (semi-simple) Lie groups.
   
Our method applies to any transformation of the
fields, even those that are not symmetries of the classical
action.  We shall develop the formalism by considering the
explicit case of an $SU(N)$ Yang-Mills theory with fermions
in the fundamental representation.  The non abelian
chiral transformation is not a symmetry of the action, and
the corresponding currents are covariantly conserved (rather
than conserved in the strict sense).  We shall compute 
their covariant divergence in the quantum theory by
following the method outlined above, and show
how the standard anomalous term \cite{NA1,NA2} arises in that approach.

\section{Compensating Fields and Conservation Laws}
\setcounter{equation}{0}

Our starting point is the Yang-Mills action
\begin{equation}
\label{LSY}
S_0 \, = \,\int d^kx \Big( - {1\over 4} Tr ( F^{\mu\nu}\,
F_{\mu\nu} ) + i\overline \psi \gamma^\mu \big(
\partial_\mu -i A_\mu \big) \psi \Big)
\end{equation}
where $A_\mu$ is a $SU(N)$-connection and the fermions are taken
in the fundamental representation.  We assume the spacetime
dimension $k$ to be even in order to have non-trivial
chirality transformations.
The action is invariant under the local transformations
\begin{equation}
\label{N3a}
\psi^\prime = \Lambda \psi, \;  
{\overline \psi}^\prime = {\overline \psi}\Lambda^{-1}, \;
A^\prime_\mu = \Lambda A_\mu \Lambda^{-1} 
\,-\, i ( \partial_\mu \Lambda ) \Lambda^{-1}
\end{equation}
where $\Lambda\,\epsilon\, SU (N)$.

The chiral infinitesimal SU(N) transformations
are
\begin{equation}
\label{CT}
\psi^\prime = \Big({\bf 1} -i \epsilon\, P_+\,\Big) \psi, \; 
{\overline \psi}^\prime = {\overline \psi}
\Big( {\bf 1}-i \epsilon  P_- \,\Big), \; 
A^\prime_\mu =  A_\mu 
\end{equation}
where 
$ P_{\pm}\,=\,{1\over 2} ( 1 \pm \,\gamma_5 )$.
Here, $\epsilon\,=\,\epsilon^a T^a\,$ is a constant element of
the  $SU(N)\,\,$ algebra.   If the connection does not vanish,
the transformations (\ref{CT}) are not symmetries of the action.
However, it is straightfoward to verify that the associated chiral
current $J_+^{\mu\,a}\,=\,{\overline \psi} \gamma_\mu T^a P_+ \psi$
is {\em covariantly} conserved,

\begin{equation}
\label{NCHI}
\Big( D_\mu J_+^\mu\,\Big)^a \,\equiv\,\partial_\mu J_+^{\mu\,a}\,
+\,f^{abc} \,A_\mu^b J_+^{\mu\,c}\,=\,0\,,
\end{equation}

\noindent where $[T^a\,,\,T^b]\,=\,if^{abc}$ defines the structure constants of the algebra.   Of course, if $A_\mu = 0$, this relation reduces to $\partial_\mu J_+^{\mu\,a}
= 0$, in agreement with the fact that the chiral transformations are then 
symmetries of the action.

It is possible to enlarge the gauge symmetry content of a field theory 
by introducing compensating fields which, as discussed in \cite{CF}, 
may lead to a different representation for the same theory, where some 
calculations become simpler.  In the present case, we will make the chiral 
transformations of (\ref{CT}), with $\epsilon = \epsilon(x)$,
become {\em gauge} symmetries of the action by introducing pure gauge
compensating group 
elements of SU(N) denoted as $g(x)$.  Being pure gauge,
the group element will have no independent equation of motion.  Rather,
its equation of motion will be precisely the covariant conservation
law (\ref{NCHI}) (at $g = {\bf 1}$).  

The constructive way to derive the action with the compensating field
included is to replace the fermionic field $\psi$
by $(P_- + P_+ g) \psi$.  If one does so, one gets the
extended action
\begin{equation}
\label{A1}
S_1\,[ \psi \, , \, {\overline \psi}\,,\,  A_\mu \, , \, g \, ] \,
=\,\int d^kx \Big( - {1\over 4} Tr ( F^{\mu\nu}\,
F_{\mu\nu} ) + i\overline \psi \gamma^\mu \big(
\partial_\mu -i {\tilde A}_\mu \big) \psi\,\Big)
\end{equation}
where ${\tilde A}_\mu$ stands for
${\tilde A}_\mu \,= {\tilde A}_\mu [ A_\mu\,,g\,]\,
= \, P_-\,A_\mu \,+\, P_+ \,B_\mu$
with
\begin{equation}
\label{bmi}
B_\mu\,=\, g^{-1} A_\mu g \,+\, i  g^{-1}\partial_\mu g \,.
\end{equation}
By construction, the action (\ref{A1}) 
is invariant under the local transformations
\begin{eqnarray}
\label{ntr}
\delta \psi &=& i (\eta(x)\,-\,\epsilon(x) P_+ \,)\psi, \; \; 
\delta {\overline \psi} = - i {\overline \psi} (\eta(x) - \epsilon(x) P_- )\no\\
\delta A_\mu &=& \partial_\mu \eta(x) \,+\, i [\eta(x)\,,\,A_\mu ], \, \,
\delta g = i ( g\epsilon(x) \,+\, [\eta(x)\,,\,g]\,)
\end{eqnarray}
which include both the original symmetry 
and the ``gauged" chiral transformations.
The complete gauge group of (\ref{A1}) is the semi-direct product
of $SU(N)$ with itself.

The theory with compensating fields is clearly classically equivalent to the
original theory.  Indeed, one can gauge away the compensating field $g$
by using the new gauge freedom.  For
instance, if we choose the gauge $g\,=\,{\bf 1}$, the
action (\ref{A1}) reduces to its original form (\ref{LSY}). 
The existence of a new gauge freedom implies further Noether
identities.  These identities relate the variational derivatives
of the action with respect to the compensating fields to the
other variational derivatives, and imply that the $g$-equations of motion
are not independent.  In fact, a straightforward calculation
yields the interesting relation
\begin{equation}
{\delta S\over \delta \beta^a}\vert_{_{\beta=0}}\,\equiv \,
-i\,\Big( \,D_\mu J_+^\mu\,\Big)^a\,,
\label{conserved}
\end{equation}
where on has set $g = {\bf 1} +\beta$ (in the vicinity of the
identity).  Thus, one can say that the compensating field
couples to the (covariant) divergence of the chiral current.  This
property will turn out to be crucial in the compution
of the quantum corrections to $\,\Big(\, D_\mu J_+^\mu\Big)^a$.

The same procedure can be followed for any group of rigid transformations
of any local action.  One may introduce the group parameters as
dynamical variables by parametrizing the fields $\phi^i$
as $\phi^i= \phi^i(g^{-1}, \phi')$ where $ \phi^i(g, \phi')$ is
the transformed of $\phi^{i'}$ under the transformation $g$.  One takes as new
variables $g$ and $\phi^{i'}$ (and drop the ').  The action is invariant under
the gauge transformations that shift in a spacetime-dependent
way the group variable $g$ by arbitrary (left) multiplication
on the group and transform
$\phi^{i'}$ accordingly.
One can use this symmetry to eliminate $g$ and recover
the original action.  In the extended formulation,
the compensating field couples
to the divergence of  the
current associated with the original rigid transformations.
Indeed, the Euler-Lagrange equations for $g$ are equivalent to 
$\partial_\mu j^\mu = 0$ if the transformations are symmetries of the 
original action, where $j^\mu$ is the Noether
current, which is conserved by Noether theorem.  This is
because the $g$'s are then ``ignorable coordinates"
of the extended action (the original action is invariant under
constant transformations and thus only derivatives of
$g$ can occur; see e.g. \cite{ZJ}).  If the original transformations
are not global symmetries of the original action,  $g$
will couple to a generalized ``covariant" divergence of $j^\mu$
rather than to the ordinary divergence, as in the specific example given
above.

\section{Quantization}
\setcounter{equation}{0}

The action (\ref{LSY}) has no gauge anomaly since there
is no chiral fermion.  The current $J_+^{\mu\,a}$ associated with the
rigid transformation (\ref{CT}) has, however, an anomalous covariant divergence.
This would seem to ruin the extended theory, since one might fear that
the new gauge symmetry will become anomalous.  If true,
equivalence with the original model would be broken at the
quantum level and potential 
inconsistencies could even arise.
That the new gauge symmetry is not afflicted by anomalies
was discussed in \cite{CF}, where it was shown that ``compensating fields
also compensate for the anomaly".  This also follows from the
general cohomological investigation of \cite{HW}.  Apart from global
cocycles related to the De Rham cohomology of the group
manifold and presumably irrelevant in perturbation theory, the
BRST cohomology group at ghost number one (related to the anomaly) has no 
cocycles involving the new ghosts.
We shall verify this property explicitly in the context
of the antifield formalism.  

To that end, we first
observe that the symmetries (\ref{ntr}) close
in an algebra,
$[\delta_1\,,\,\delta_2\,]\,\phi^i\,=\,\delta_3\phi^i$
for any field $\phi^i\,$. The parameters of the transformation 
on the right hand side are given by
$\eta_3 = i [\eta_1\,,\,\eta_2\,]$ and 
$\epsilon_3 = i \Big( [\eta_1\,,\,\epsilon_2]\,
+\,[\epsilon_1\,,\,\eta_2]\,-\,[\epsilon_1\,,\,\epsilon_2]\,\Big)$.
The BV action (at zero order in $\hbar$) 
then follows by the standard procedure,
\begin{eqnarray}
\label{BVA}
S &=& S_1 + \int d^kx \Big( i \psi^\ast ( c -b P_+)
\psi-i{\overline \psi}
(c -b P_-){\overline \psi}^\ast
+ Tr\{ i g^\ast ( g\,b
+\, [c\,,\,g]) \nonumber \\ 
&+&  A^\ast_\mu D^\mu c\,+\, {i\over 2} c^\ast [c\,,\,c]\,
-\,
{i\over 2} b^\ast ( [b\,,\, b ]\,-2 [c\,,\,b] )\,\}\,\, \Big)
\end{eqnarray}
where we have introduced the ghosts $c$ and $b$ corresponding to 
the parameters $\eta$ and $\epsilon$ respectively and also 
the antifields associated with each field. 

We will  represent the total sets of fields and antifields 
respectively as $\,\{\varphi^I\}\,$,
$\,\{\varphi^{\ast}_I\}\,$.
Defining the BRST transformation of any quantity $X$ as: 
$\,\,sX\,=\,(X\,,\,S)$, where $(X,Y)$  
is the standard antibracket \cite{BV1}, it is not difficult to see that 
$s^2 X\,=\,0$ and that the classical master 
equation $(S\,,\,S)\,=\,0$ is valid.
The BRST transformations of the fields are
given by (\ref{ntr}) with the gauge parameters replaced by
the ghosts, as well as
$s c = i\, c\,c$,  
$s b = - i \, b\,b \,+\,i\, [\,c\,,\,b\,]$.

It is useful to introduce the invariant forms
$\sigma=-ig^{-1}sg
=b-c+g^{-1}cg$.
These
fulfill the Maurer-Cartan Eqs.
$s\sigma=i\sigma^2$.
The related form $\sigma_\mu=
-ig^{-1}\partial_\mu g$
transforms as
$s\sigma_\mu=\partial_\mu\sigma-i[\sigma,\sigma_\mu]$.
Furthermore,
$s(g^{-1}A_\mu g)=i[c-b,g^{-1}A_\mu\, g]+g^{-1}\partial_\mu c\, g$
and as $B_\mu=-\sigma_\mu+g^{-1}A_\mu g$, 
one gets
$s B_\mu\,=\, \partial_\mu(c-b)+\,i\,[\,c-b,B_\mu\,]$.
This equation shows explicitly that the ghost
associated with the connection $B_\mu$ is $c-b$.

\section{One-loop order master equation}
\setcounter{equation}{0}

The BV vacuum functional is defined as 
\begin{equation}
Z_{\overline \Psi} \,=\,\int [d\varphi^I] \, \delta [ \varphi^\ast_I 
- {\partial \overline\Psi\over \partial \varphi^I} ] exp ( {i\over \hbar } W )
\end{equation}
with $W\,=\,S\,+\,\hbar M_1\,$.  Properly speaking, one should include
the non-minimal sector.  This will be done below, but since these variables
do not affect the cohomological considerations (they form trivial pairs
\cite{HT}), they will not be written explicitly here.
The BV vacuum functional is independent of the choice of gauge fermion
${\overline \Psi}$  if, besides the classical master equation, 
the one loop order master equation is also satisfied
\begin{equation}
\label{Master1}
(M_1,S) \,= \, \,i\, \Delta S\,\,,
\end{equation}
where $\Delta \equiv
{\delta_r\over\delta\phi^A}{\delta_l\over\delta\phi^\ast_A}\;$.
This equation is undefined unless we regularize the action 
of the $\Delta$ operator. 
Using a Pauli Villars (PV) regularization, 
with usual mass terms for the PV fermionic fields, the 
four dimensional case ($k=4$) to which we shall restrict
our attention from now on, can be written in the form \cite{TPN}
\begin{equation}
\label{14}
\Delta S \,=\, \alpha \,tr\int d^4x \, \left( \,\partial_\mu \,(c-b)\,
\Delta^\mu_B-\partial_\mu c \, \Delta^\mu_A \,\right)\,.
\end{equation}
Here,
\begin{equation}
\Delta^\mu_A=\epsilon^{\mu\nu\rho\sigma}(A_\nu\partial_\rho A_\sigma-
{i\over2}A_\nu A_\rho A_\sigma)
\label{13}
\end{equation}
and $\Delta^\mu_B$ is given by a similar expression, with the
replacement $A_\mu\rightarrow B_\mu\,$ given by (\ref{bmi}).  
In the above expression,  $\alpha\,=\, -\,{1\over 24 \pi^2}\,$. 
We are assuming that the measure for the $g$ sector is BRST invariant
(we are taking the Haar measure).

The term $tr\int d^4x \partial_\mu c \Delta^\mu_A$ is the
standard ABBJ (Adler-Bardeen-Bell-Jackiw) anomaly for
the  gauge field $A_\mu$.  It can be rewritten in
form notations as $tr \int dc(A dA - (i/2)A^3) \equiv
\int a^{ABBJ}_A$ and
is well known to be a solution
of the Wess-Zumino consistency condition.
Similarly, the term $tr\int d^4x \partial_\mu (c-b) \Delta^\mu_B$,
which can be rewritten $tr \int d(c-b) (B dB - (i/2)B^3)\equiv
\int a^{ABBJ}_B$
is the ABBJ anomaly with $B_\mu$ instead of $A_\mu$.
It also solves the Wess-Zumino consistency condition.
Consequently, for the full $\Delta S$, one has
$s\,\Delta S\,=\,( \Delta S\,,\,S)\,=\,0$.

The quantity $\Delta S$ represents 
the variation of the
path integral measure under BRST transformations.  
If this variation cannot be
compensated by the variation of some local counterterm then the 
theory would be anomalous, and this would be a priori a disaster
since it is a gauge anomaly.
So, the important point now is to find out if there is a 
local counterterm $M_1$, to be
added to the action, whose BRST variation cancels the 
candidate anomaly $\Delta S$.  It was shown in \cite{AB} that
such a counterterm exists in the Abelian case.  We extend
this result here to the non-abelian case.

It is rather easy to see that $\Delta S$ is $s$-exact
in the space of local functionals, and thus that $M_1$ exists.
Indeed, it is well known that the ABBJ anomaly is related to the
invariant $tr F^3$ in 2 dimensions higher, i.e., here, in six
dimensions through a chain of descent equations (see e.g. \cite{zumi}).
Explicitly, one has
$tr F^3_A = dQ^{5,0}_A$ where $Q^{5,0}_A$ is the Chern-Simons
$5$-form constructed out of $A$, and $sQ^{5,0}_A +
dQ^{4,1}_A = 0$, where $Q^{4,1}_A $ is the ABBJ anomaly $a^{ABBJ}_A$.
In $Q^{i,j}$, the first superscript is the form-degree, while the
second superscript is the ghost number.
Similarly, one gets $tr F^3_B = dQ^{5,0}_B$ where $Q^{5,0}_B$ 
is the Chern-Simons $5$-form constructed out of $B$, and $sQ^{5,0}_B+
dQ^{4,1}_B = 0$, where $Q^{4,1}_B $ is the ABBJ anomaly $a^{ABBJ}_B$.
Now, because $A$ and $B$ are related by a gauge transformation
(Eq. (\ref{bmi})), they have field strengths related as
$F_B = g^{-1} F_A g$ and thus $tr F^3_A = tr F^3_B$.
This implies $d(Q^{5,0}_B - Q^{5,0}_A) = 0$, i.e.,
$Q^{5,0}_B - Q^{5,0}_A = dM^{4,0}$ for some $4$-form
$M^{4,0}$.  Substituting this result in the next descent equation 
yields then $d(Q^{4,1}_B - Q^{4,1}_A- sM^{4,0})= 0$,
i.e., 
\begin{equation}
a^{ABBJ}_B - a^{ABBJ}_A \equiv Q^{4,1}_B - Q^{4,1}_A
= s M^{4,0} + d M^{3,1}
\label{ABBJ}
\end{equation}
for some $3$-form $M^{3,1}$.
This implies, upon integration, that $\Delta S$ is
indeed exact,
\begin{equation}
i \Delta S = i \alpha \int (a^{ABBJ}_B - a^{ABBJ}_A)
= s M_1 , \; \; M_1 = i \alpha \int M^{4,0}.
\end{equation}

The explicit form of the counterterm $M_1$ may be found either by
following the above procedure, or by using a perturbative
expansion in the number of fields.  One gets

\begin{eqnarray}
M_1 &=& \alpha \big(\int_{\cal M} d^5x  
\epsilon^{\mu\nu\rho\omega\lambda} 
({i \over 10}
Tr ( \sigma_\mu \sigma_\nu \sigma_\rho \sigma_\omega \sigma_\lambda\,)\,
\no \\ &+& \,\int_{\partial {\cal M}} \,d^4x\,\epsilon^{\mu\nu\rho\sigma}\,Tr \,\Big\lbrace \, 
\partial_\mu g \,g^{-1}
\Big( A_\nu \partial_\rho  A_\sigma -\,{i\over 2}
A_\nu A_\rho A_\sigma  \Big) \no\\ &-&
{1\over 4}\,\partial_\mu g  g^{-1}
A_\nu  \partial_\rho g g^{-1}  A_\sigma  + {1\over 2}
\partial_\mu g g^{-1} \partial_\nu g g^{-1} A_\rho  A_\sigma 
\no\\ &-& {i\over 2} \partial_\mu g g^{-1} \partial_\nu g g^{-1}
\partial_\rho g g^{-1} A_\sigma
\Big\rbrace) \label{Counter}
\end{eqnarray}

\noindent where the first term is the Wess-Zumino term which, as usual, 
is defined on an $k+1$ dimensional manifold  ${\cal M}$, with 
boundary $\partial {\cal M}\,$ given by the four-dimensional space on which the theory is defined.

\section{Anomalous divergence}
\setcounter{equation}{0}

We will now show how to get the anomalous divergence of the 
chiral current from the previous results.
The quantum BV action $ W $ is the sum of the counterterm from 
the last section with the BV classical action of equation (\ref{BVA}). 
Let us introduce also a trivial pair of fields ${\bar \pi}, {\bar b}$, 
and the corresponding antifields  ${\bar \pi}^\ast , {\bar b}^\ast $, 
in order to allow an appropriate gauge fixing of the extra symmetry. We
should also introduce non minimal variables for the original
gauge symmetry, but these will not be written explicitly.  So we have
$W \,=\, S \,+\, \hbar M_1\,+\, \int d^4x Tr ( {\bar \pi}  {\bar b}^\ast )$.
We choose  a gauge fixing fermion of the form
$\overline \Psi\,=\,Tr \Big( {\bar b} ( g\,-\,{\bf 1}\,-\,\beta\,)\,\Big)\,+\, 
\Psi$ 
where $\beta$ is an (infinitesimal)
external function and where $\Psi$ does not depend
on the fermionic variables or the extra fields $g,b \,,\bar{\pi}$ and $ \bar{b}$.
This choice of gauge fixing fermion enforces the gauge $g={\bf 1} +
\beta$ and from what we have seen above, $\beta$ will appear as a source
for the covariant divergence of the chiral current.

A direct calculation gives as gauge fixed quantum action  
\begin{eqnarray}
W_{_\Sigma} &=& W [\phi^I,\,\varphi^\ast_I \,= 
{\partial \overline\Psi\over \partial \varphi^I}\,]\, 
=\,\int d^4x \Big( - {1\over 4} Tr ( F^{\mu\nu}\,
F_{\mu\nu} ) \no \\
&+& i\overline \psi \gamma^\mu \big(
\partial_\mu -i {\tilde A}_\mu \big) \psi 
+  Tr\, \{ i {\bar b} ( g\,b\,+\, [c\,,\,g] )
+ {\partial \Psi\over \partial A^\mu} D^\mu c\, \no \\
&+&\, {i\over 2} {\partial \Psi \over \partial c} [c\,,\,c]\,
-\,{\bar \pi}  ( g\,-{\bf 1}\,-\,\beta )\,  \}\,+\, \hbar M_1\,\Big)
\end{eqnarray}
Now we build up the vacuum functional 
$Z \,[\beta ]\,=\,\int [d\varphi^I]\,exp \{\,{ i\over \hbar}\, W_\Sigma \,\}$
where we are omitting the dependence on $\Psi$ in the notation. 
Then we integrate over the fields ${\bar \pi}$ and $g$. This 
amounts to substituting $g$ by  ${\bf 1}\,+\,\beta$. 
The fermionic term of the action becomes
\begin{equation}
\label{N1}
i\overline \psi \gamma^\mu \big(
\partial_\mu -i A_\mu \big) \psi\,+\, 
\overline \psi \gamma^\mu P_+\,(\,
i \partial_\mu \beta \,+\,  [ A_\mu\,,\,\beta\,] \,)\,\psi. 
\end{equation}
The integration over $b$ and ${\bar b}$ is direct and
yields one together with the Haar measure.
The vacuum functional becomes therefore
\begin{equation}
Z\,[\beta ]\,=\,\int [d\phi^A]\,exp \{\,{ i\over \hbar}\, 
{\overline W}_\Sigma \,\}
\end{equation}
where $\{\phi^A\}\,=\,\{A_\mu , c , ...\}$ (the dots refer to 
trivial pairs associated to the original gauge symmetry)
and where ${\overline W}_\Sigma$ is just the sum of the (extended)
classical action and the counterterm of eq. (\ref{Counter}) 
in the gauge $g = {\bf 1} + \beta$,
plus non-minimal terms that do not involve $\beta$.
As the master equation is satisfied, this object is independent 
of the external parameter    $\beta$ (``Fradkin-Vilkovisky
theorem"). Thus we have

\begin{equation}
{\delta Z[\beta]\over \delta \beta^a}\vert_{_{\beta = 0}}\,
%=\, \int [d\phi^A] {i\over\hbar}\,\Big(\,{\delta {\overline W}_\Sigma 
%\over \delta \beta^a}\,exp \{\,{ i\over \hbar}\, {\overline W}_\Sigma \,\}
%\Big)\vert_{_{\beta = 0}}\,
=\,
{i\over \hbar }\, \langle\, {\delta {\overline W}_\Sigma 
\over \delta \beta^a}\,\rangle\vert_{_{\beta = 0}}\,\,=\,0\,\,.
\end{equation}

{}From the equation (\ref{Counter}) we find
\begin{equation}
{\delta M_1 \over \delta \beta^a}\vert_{_{\beta = 0}}\,=\, -\,\,
\alpha  \epsilon^{\mu\nu\rho\sigma} Tr \,\lbrace \, 
 \partial_\mu 
\Big( A_\nu \partial_\rho  A_\sigma \,-\,{i\over 2}
A_\nu A_\rho A_\sigma  \Big) T^a \rbrace\,\,.
\end{equation}
\noindent Using this equation as well as (\ref{N1}) and 
(\ref{conserved}) 
we finally get

\begin{eqnarray}
\hbar\,{\delta Z[\beta]\over \delta \beta^a}\vert_{_{\beta = 0}} &=& \langle\, 
\Big( D_\mu J_+^\mu\,\Big)^a
 -\,i\hbar\alpha \epsilon^{\mu\nu\rho\sigma}\, 
Tr \,\lbrace \, 
 \partial_\mu 
\Big( A_\nu \partial_\rho  A_\sigma \,-\,{i\over 2}
A_\nu A_\rho A_\sigma  \Big) T^a \rbrace\,\rangle \no\\
&=& 0
\end{eqnarray}
where now the expectation values are taken in the original theory, 
with no extra variables.
This reproduces the desired anomalous divergence.

\section{Conclusion}
\setcounter{equation}{0}

We have shown that anomalous expectation values of currents associated 
with global transformations can be calculated by introducing compensating fields.
We have performed the analysis in the antifield formalism.
The present work extends the previous Abelian study of \cite{AB}. 
We have shown that the new gauge symmetries associated with the compensating fields  
are not obstructed at the quantum level since they do not change the 
cohomology of the theory.
However, it is necessary to introduce quantum corrections to the BV action in 
order to fulfill the quantum master equation. These quantum corrections 
precisely generate the appropriate anomalous contribution to the
divergence of the global current, although no gauge anomalies are present.
Our procedure represents thus an interesting example where quantum corrections 
have a non trivial role.

\vskip 1cm
\noindent {\bf Acknowledgment:} This work is partially supported  by
 CNPq, FINEP and FUJB (Brazilian Research Agencies). M. H. is grateful
 to Glenn Barnich for useful comments, as well as to CNPq and
 the physics departments of UERJ and UFRJ for kind hospitality.

\end{document}